\documentclass[12pt]{article}
\usepackage{amssymb}

\def\del{\partial}
\def\Ds{D\!\!\!\! /}

\def\B{\mathcal{B}}
\def\X{\mathcal{X}}

\def\vD{\vec{D}}
\def\vt{\vec{\tau}}
\def\kok{\sqrt{2}}
\def\yarim{{{1}\over{2}}}

\def\a{\alpha}
\def\b{\beta}

\def\d{\delta}
\def\e{\epsilon}
\def\se{\varepsilon}

\def\la{\lambda}

\def\s{\sigma}

\def\f{\phi}

\def\x{\xi}

\def\adot{\dot{\alpha}}
\def\bdot{\dot{\beta}}

\def\cbar{\bar{c}}

\def\labar{\bar{\lambda}}
\def\sbar{\bar{\sigma}}

\def\xbar{\bar{\xi}}

\def\bs{\bar{s}}
\def\ft{\phi^{\dag}}

\def\se{\mathcal{E}}

\def\ts{\tilde{s}}

\def\be{\begin{equation}}
\def\ee{\end{equation}}
\def\bea{\begin{eqnarray}}
\def\eea{\end{eqnarray}}

\def\kok{\sqrt{2}}
\def\yarim{{{1}\over{2}}}

\begin{document}

%

\thispagestyle{empty}
\begin{flushright}hep-th/0304154\\
\end{flushright}
\vskip5em
\begin{center}

{\Large{\bf $N=2$ super Yang-Mills action as a }}\\

{\Large {\bf Becchi-Rouet-Stora-Tyutin term,  }}\\

{\Large {\bf topological Yang-Mills action and instantons}}
\vskip1cm
K. \"{U}lker
\vskip2em
{\sl Feza G\"{u}rsey Institute,}\\
{\sl \c{C}engelk\"{o}y, 81220, \.{I}stanbul, Turkey}\\
\vskip1.5em
\end{center}
\vskip1.5cm
%
\begin{abstract}

\noindent By constructing a nilpotent  extended BRST operator $\bs$ that involves the N=2 global supersymmetry
transformations of one chirality, we show that the standard N=2 off-shell Super Yang Mills Action can be represented as
an exact BRST term $\bs \Psi$, if the gauge fermion $\Psi$ is allowed to depend on the inverse powers of supersymmetry
ghosts. By using this nonanalytical structure of the gauge fermion (via inverse powers of supersymmetry ghosts), we give
field redefinitions in terms of composite fields of supersymmetry ghosts and N=2 fields and we show that Witten's
topological Yang Mills theory can be obtained from the ordinary Euclidean N=2 Super Yang Mills  theory directly by using
such field redefinitions. In other words, TYM theory is obtained as a change of variables (without twisting).  As a
consequence it is found that physical and topological interpretations of N=2 SYM are intertwined together due to the
requirement of analyticity of global SUSY ghosts. Moreover, when after an instanton inspired truncation of the model is
used, we show that the given field redefinitions yield the  Baulieu-Singer formulation of Topological Yang Mills. 

\end{abstract}
\vskip4em
PACS codes: 11.30.Pb, 11.27.+d, 12.60.Jv\\
Keywords : Supersymmetric gauge theories, BRST symmetry, Topological theories, Instantons
\vskip2em
{\small{e-mail: kulker@gursey.gov.tr}}
\vfill
\eject
\setcounter{page}{1}
%
%
\section{Introduction}

N=2 super Yang - Mills (SYM) theory has been extensively studied these last years after the work of Seiberg and
Witten \cite{sw} where a self consistent non-perturbative effective action was calculated
by using certain ans\"{a}tze dictated by physical intuitions. This solution is unique \cite{fmrss}.

After this seminal paper \cite{sw} one of the main area of research in N=2 SYM is to calculate directly the
multi-instanton contributions to the holomorphic prepotential and consequently to check the correctness of the results
of \cite{sw}. These multi-instanton contributions are calculated in a pioneering work \cite{dkm} for one and two
instantons by using a semi-classical expansion around {\it approximate} saddle points of the action \cite{dkm}
\footnote{One instanton contribution is also calculated in \cite{finnel, ito}.}. The results are found to be in
agreement with the one of \cite{sw}. (For a self-contained review see Ref. \cite{dkhm}.) However a natural question was
posed by several researchers: how it could be possible that an approximate approach gives an exact result and what was
the mechanism behind \cite{fb}?

In order to get answer to the above questions, in Ref.\cite{bftt2} the instanton calculus is performed in the framework
of Topological Yang Mills (TYM) theory and it is found that the results are the same with that of {\it traditional}
instanton calculus of \cite{dkm}. As noted in \cite{bftt2}, the underlying fact of this result is, the action of TYM
theory, which can be obtained as a twist of the ordinary N=2 SYM \cite{witten},
can be written as a BRST-exact term \cite{bs} and the functional integration over the antifields of the topological
theory gives the same field configurations of the constrained instantons of the ordinary theory without any
approximation. Therefore, the authors of Ref. \cite{bftt2} have noted that the twisting procedure can be thought as
variable redefinitions in flat space time.

However, up to now no explicit variable redefinitions is given in order to obtain the topological theory. For instance,
even if the twist can be considered as a linear change of variables, after twisting physical interpretation of some
fields change, i.e. some become ghosts or anti-ghosts.   The dimensions of the topological fields are also
different then their untwisted counterparts.

Our main aim in this note, is to derive explicit field redefinitions in order to obtain the topological fields with
correct dimensionality  and ghost number.The strategy that we follow is to use the the extension of BRST formalism (also
called BV or field-antifield formalism \cite{bv,gps}) to include global supersymmetry (SUSY). \cite{w,mag,mpw}
\footnote{Such an extension of BRST transformations that includes rigid symmetries is first introduced in \cite{bon}.
The problem of how to extend the BRST formalism to include arbitrary global symmetries can be found in Ref.
\cite{bra1}.}. By using this formalism we show that N=2 SYM action can be written as an exact term and both of the
formulations of TYM given by Witten \cite{witten} and Baulieu and Singer \cite{bs} can be obtained as a change of
variables without twisting.

The paper is organized as follows: In Sec.II, we review briefly how to include SUSY in an extended BRST operator. Since
the actions of SYM theories can be represented as chiral (or antichiral) multiple
supervariations of lower dimensional gauge invariant field polynomials \cite{ben1, ben2}\footnote{For a similar approach
of constructing N=1 globally and locally supersymmetric actions and also for the discussion of anomalies, see
\cite{bra2}.}, in Sec.III we construct a nilpotent extended BRST operator $\bs$ on the field space
that only contains the chiral part of the N=2 SUSY as a global symmetry in addition to gauge symmetries.  It is then
straightforward to find an expression as a {\it gauge fermion} of which the full off-shell N=2 action is its BRST
($\bs$) variation under some conditions in Minkowski space. In Sec. IV, we give Euclidean formulation of the result
given in Sec.III in order to derive field redefinitions  by comparing the Euclidean N=2 SYM action written as an
exact term and the topological theory. We demonstrate explicitly that Wittens topological Yang-Mills theory can be
obtained from the ordinary Euclidean N=2 SYM directly by using such field redefinitions. In Sec.V, by using an instanton
inspired truncation of the model given by Zumino \cite{zumino}, we also show how to obtain the approach of
Baulieu-Singer of TYM with the help of field redefinitions. Finally, Sec.VI is devoted to concluding remarks.

\section{N=2 SYM and Extended BRST  transformations }

N=2 SYM theory is a rather old and well studied theory. In this work we will study the off-shell formulation of the
theory given in \cite{gsw,soh} by using the conventions of \cite{wb}. The action of the theory is given in Minkowski
space as
\bea
S_{N=2} &=& \frac{1}{g^2}Tr \int d^4 x (-\frac{1}{4} F_{\mu\nu} F^{\mu\nu}  -i\la^i D \!\!\!\! / \labar_i  + \f D_{\mu}
D^{\mu} \f^{\dag}\nonumber\\
&&\qquad\qquad\qquad  -\frac{i\kok}{2}(\la_i [\la^i , \ft ] +\labar^i [\labar_i ,\f ]) -\yarim [\f ,\ft]^2 +
\yarim\vD .\vD)
\eea
where the gauge field $A_\mu  $ and the scalar fields $\f\, ,\,\ft $ are singlets, the Weyl spinors $\la_{i\a} \,\,
\labar^{i}_{\adot}$ are doublets and the auxiliary field $\vD $ is a triplet under the $SU(2)_R$ symmetry group. These
fields are members of N=2 vector multiplet $V=(A_\mu\, ,\,\f \, ,\, \ft \, ,\, \la_{i\a} \, ,\,
\labar^{i}_{\adot} \, ,\, \vD )$ \cite{gsw,soh}. The  $SU(2)_R$ indices of the spinors are raised and lowered due to
\bea
\la^i = \se^{ij} \la_j \quad&,&\quad \la_i = \la^j \se_{ji}
\\
\labar_i = \se_{ij} \labar^j \quad&,&\quad \labar^i = \labar_j \se^{ji}
\eea
where the antisymmetric tensor $\se^{ij}$ is given as\footnote{Note that in our convention the $\se^{ij}$ is different
then the one, $\e_{\a\b}$,  used for spinor indices},
$$
\se_{12} = \se^{12} = -\se_{21} = - \se^{21}=1.
$$

The extension of the BRST transformations with the global N=2 SUSY and translation symmetry is first given in
Ref.\cite{mag}:
\be
s=s_0 -i \x^i Q_i -i \xbar_i \bar{Q}^i -i \eta^{\mu}\del_{\mu}
\ee
where $s_0$ is the ordinary BRST transformations, $Q_i \, ,\, \bar{Q}^i $ are chiral and antichiral parts of N=2 SUSY
transformations and $\x^{i\a} \, ,\, \xbar_{i\adot} $ and $\eta _{\mu}$ are the constant commuting chiral,
antichiral SUSY ghosts and constant imaginary anticommuting translation ghost respectively. Note that the parameters of
the global transformations are promoted to the status of constant ghosts and their Grassmann parity are changed so that
the  extended transformation $s$ is a homogeneous transformation.

The extended BRST transformations read on the elements of the N=2 vector multiplet as\footnote{Here, $\vt$'s are Pauli
spin matrices},
\bea
sA_{\mu}&=& D_{\mu}c +\x_i \s_{\mu}\labar^i + \xbar^i \sbar_{\mu}\la_i -i \eta^{\nu}\del_{\nu}A_{\mu}\\
s\la_i &=& i\{c,\la_i \} - i\s^{\mu\nu}\x_i F_{\mu\nu} +\x_i [\f ,\ft ]  -\kok \s^{\mu}\xbar_i D_{\mu} \f + \vt_i ^j
\x_j.\vD -i \eta^{\mu}\del_{\mu} \la_i\\
s\labar^i &=& i \{ c,\labar^i \} - i \sbar^{\mu\nu} \xbar^i F_{\mu\nu} -\xbar^i [\f ,\ft ]  -\kok\sbar^{\mu}\x^i D_{\mu}
\ft -\xbar^j \vt_i ^j . \vD -i \eta^{\mu}\del_{\mu}\labar^i  \\
s\f &=& i[c,\f ]-i\kok \x_i\la^i -i \eta^{\mu}\del_{\mu}\f\\
s\ft &=& i[c,\ft ]-i\kok \xbar^i \labar_i -i \eta^{\mu}\del_{\mu}\ft\\
s\vD &=& i[c, \vD] + i\vt_i ^j(\x_j \Ds\labar^i -\xbar^i\bar{\Ds}\la_j  +\kok\x^i [\la_j ,\ft] -\kok\xbar_j
[\labar^i,\f]) -i \eta^{\mu}\del_{\mu}\vD
\eea
On the other hand, in order to get a nilpotent $s$,
\be
s^2 = 0
\ee
 Fadeev-Popov ghost field and the global ghosts are asked to transform as
\bea
sc &=& \frac{i}{2}\{ c,c\}-2i\x_i\s^{\mu}\xbar^i A_{\mu} -\kok\x_i\x^i \ft -\kok\xbar^i\xbar_i \f -i
\eta^{\mu}\del_{\mu}c \\
s\eta_{\mu}  &=& - 2\x^i\s_{\mu}\xbar_i  \\
s\x_i &=& s\xbar_i = 0
\eea

Note that with the help of extra terms in $sc$, the characteristic complication that SUSY algebra is modified by
field-dependent gauge transformations is solved, whereas the closure on translations disappeared due to inclusion of
translation ghosts. We summarize the dimension, ghost number and the R-charges of the fields and ghosts in Table 1.
%
\begin{table}[hbt]
\centering
\begin{tabular}{|c||c|c|c|c|c|c|c|c|c|c|}
\hline
&$A_\mu$&$\f$&$\ft$&$\la^i$&$\labar_i$&$\vD$&$\,\,c\,\,$&$\x^i$&$\xbar_i$&$\eta_{\mu}$\\
\hline
$Dim$&1&1&1&3/2&3/2&2&0&-1/2&-1/2&-1 \\
\hline
$Gh$&0&0&0&0&0&0&1&1&1&1 \\
\hline
$R$&0&-2&2&-1&1&0&0&-1&1&0\\
\hline
$GP$&0&0&0&1&1&0&1&0&0&1 \\
\hline
\end{tabular}
\caption[t1]{Dimensions   $d$, Grassmann parity $GP$, ghost number $Gh$
  and R-weights.}
\end{table}
\section{N=2 SYM action as an exact term}

It is known from cohomological arguments that the actions of SYM theories can be represented as chiral (or antichiral)
multiple supervariations of lower dimensional gauge invariant field polynomials \cite{ben1, ben2}. For instance,
  N=2 SYM action can be written as a fourfold chiral SUSY transformation of  $Tr\f ^2$  in component formalism of
SUSY.

On the other hand, from the definition of $s$ it is still possible to derive another nilpotent operator by using a
suitable filtration of global ghosts. We choose this filtration to be
\be
\mathcal{N} = \xbar_{i\adot} \frac{\d}{\d \xbar_{i\adot}} + \eta_{\mu} \frac{\d}{\d \eta_{\mu}} \quad ;\quad s =
\sum{s^{(n)}} \quad ,\quad [\mathcal{N},
s^{(n)}]=ns^{(n)},
\ee
so that the zeroth order in the above expansion is a operator that includes ordinary BRST and chiral SUSY on the space
of the fields of the N=2 vector multiplet
\be
\bs:= s^{(0)} = s_0 -i \x^i Q_i
\ee
that is also nilpotent,
\be
\bs ^2 =0.
\ee
The $\bs$ transformation of the fields are now given as
\bea
\bs A_{\mu}&=& D_{\mu}c +\x_i \s_{\mu}\labar^i \\
\bs \la_i &=& i\{c,\la_i \} - i\s^{\mu\nu}\x_i F_{\mu\nu} +\x_i [\f ,\ft ] + \vt_i ^j \x_j.\vD \\
\bs \labar^i &=& i \{ c,\labar^i \}  -\kok \sbar^{\mu}\x^i D_{\mu} \ft \\
\bs \f &=& i[c,\f ]-i\kok \x_i\la^i \\
\bs \ft &=& i[c,\ft ]\\
\bs \vD &=& i[c, \vD] + i\vt_i ^j(\x_j \Ds\labar^i  + \kok\x^i [\la_j ,\ft]) \\
\bs c &=& \frac{i}{2}\{ c,c\} -\kok\x_i\x^i \ft\\
\bs \eta_{\mu}  &=& \bs \x_i = \bs \xbar_i = 0
\eea

Since $\bs$ contains gauge transformations and chiral SUSY and the action is gauge invariant, it is straightforward to
assume that the action can be written also as an $\bs$ exact term of a gauge invariant field polynomial which is
independent of Fadeev-Popov ghost fields\footnote{In other words, we assume that the action can be chosen to be a
trivial element of equivariant cohomology of $\bs$.  See for instance Ref.s\cite{sorlec,sor1,sor2} and the references
therein.},
\be
I=\bs \Psi.
\ee

It is clear that $\Psi$, the so called gauge fermion in BV formalism, has negative ghost number, $Gh(\Psi )=-1$.
However, since no fields with negative ghost number has been introduced and since we have chosen the gauge fermion to be
free of Fadeev-Popov ghosts, the only way to assign a negative ghost number to $\Psi$ is to choose $\Psi$ to depend on
the negative powers of the global SUSY ghosts. In other words, the action can be written as an exact $\bs$-term only if
the extended BRST operator $\bs$ is defined on the space of field polynomials that are not necessarily analytic in
constant ghosts.

Therefore a further assumption to assign a negative ghost number to $\Psi$  is to choose a gauge fermion that
has the following form:
\be
\Psi=\frac{1}{\x _k \x^k} \x^i \int d^4 x \psi_i
\ee
where $\psi_i ^\a $ is a dimension $7/2$  fermion that is made from the fields of the N=2 vector multiplet. The most
general such gauge fermion that is covariant in its Lorentz, spinor and $SU(2)_R$ indices is easy to find:
\be
\Psi=\frac{1}{\x _k \x^k}Tr \int d^4 x \{ (k_1 \x ^i  \la_i  [\f ,\ft ] + k_2 \x^i \vt _i ^j \la_j . \vD+ k_3 \x^i
\s^{\mu\nu}\la_i F_{\mu\nu} + k_4 \f \x^i \Ds \labar_i ) \}
\ee
Here $k$'s are constants. In order that the $\bs$ variation of $\Psi$ to be free of chiral ghosts after some algebra it
is seen that constants $k$ should be fixed  and as a result
\bea
\bs\Psi &=& \bs \frac{1}{\x _k \x^k} Tr \int d^4 x \{ (\yarim \x ^i  \la_i  [\f ,\ft ] -\yarim \x^i \vt _i ^j \la_j .
\vD  -\frac{i}{2} \x^i \s^{\mu\nu}\la_i F_{\mu\nu} +\frac{\kok}{2} \f \x^i \Ds \labar_i ) \}\\
&=&Tr \int d^4 x (-\frac{1}{4} F_{\mu\nu} F^{\mu\nu}
-\frac{i}{8}\e^{\mu\nu\la\rho}F_{\mu\nu}F_{\la\rho}  -i\la^i D \!\!\!\! / \labar_i  + \f D_{\mu} D^{\mu}
\f^{\dag}\nonumber\\
&&\qquad\qquad\qquad -\frac{i\kok}{2}(\la_i [\la^i , \ft ] +\labar^i [\labar_i ,\f ]) -\yarim [\f ,\ft]^2 + \yarim\vD .
\vD)
\eea
is found. This expression is exactly the action (1) of N=2 SYM theory with a topological term
$\e^{\mu\nu\la\rho}F_{\mu\nu}F_{\la\rho} $ ,
\be
I_{N=2} = \frac{1}{4\pi} Im [\Upsilon \, (\bs\Psi)] = S_{N=2}- \frac{\theta}{16\pi}Tr \int d^4 x
\e^{\mu\nu\la\rho}F_{\mu\nu}F_{\la\rho}
\ee
where $\Upsilon$ is the complex coupling constant, $ \Upsilon = \frac{i4\pi}{g^2}+\frac{\theta}{2\pi}$.

One can argue that the action written in this form is obvious since the action can be obtained by applying all four
supersymmetry transformations to the prepotential $Tr\f^2$. However, it has important consequences.

First of all, the operator $\bs$ is strictly nilpotent. It is a well known fact that the cohomology of the complete
operator $s$, given  in (4), is isomorphic to a subspace of the cohomology of $\bs$, since $\bs$ is obtained after using
a suitable filtration of $s$ (see for instance \cite{psbook,sorlec}). If the functional space is defined to be the
polynomials of the fields that are not necessarily analytic in the constant ghosts $\x_i$, as it is shown above, the
action belongs to the trivial cohomology of $\bs$ and therefore to that of complete operator $s$.

Second, the above given nonanalyticity argument plays an important role to obtain the topological Yang-Mills from
Euclidean N=2 SYM by identifying the topological fields with certain functions of fields and SUSY ghosts of N=2 SYM
(which are also partly nonanalytic $\x_i$'s). These points will be clarified in the next sections.

\section{Euclidean N=2 SYM and\\TYM as a variable redefinition }

As it is well known TYM theory can be obtained by twisting N=2 SYM theory in Euclidean space \cite{witten}. In summary
twisting procedure is simply identifying the $SU(2)_R$ index $i$ with the spinor index of one
chirality  i.e. $\a$ \footnote{In $\mathbb{R}^4$ the symmetry group of N=2 SYM is $SU(2)_L \otimes SU(2)_R \otimes
SU(2)_R \otimes U(1)_R $. The twist $( i\equiv\a )$ consists of replacing the rotation group $SU(2)_L \otimes SU(2)_R $
with $SU(2)_L ' \otimes SU(2)_R$ where $SU(2)_L '$ is the diagonal sum of $SU(2)_L \otimes SU(2)_R $. For a detailed
analysis of topological theories see for instance \cite{top}.}  and $R$-charges of the fields with ghost number. It is
then possible to show that TYM action can be written as a BRST-exact term up to some field redefinitions \cite{bs}.
Therefore  TYM theory is called a topological theory of cohomological type and  it is natural to look for an
analogy between the results of previous section and TYM theory by rewriting the results of the previous section in
Euclidean space.

However, before formulating the results of  Sec.III in Euclidean space, we find it useful to clarify our
approach to Euclidean N=2 SYM. First of all, obviously in Euclidean space the chiral and antichiral spinors are not
related with each other. Nevertheless, it is still possible to find consistent reality conditions for the spinors of
extended supersymmetry \cite{bvn}. We will take these reality conditions in our conventions as,
\be
(\la_i ^\a )^\dag = i \e_{\a\b} \se^{ij} \la_j ^\b \quad ,\quad (\labar^{i\adot})^\dag =i \e_{\adot\bdot}
\se_{ij}\labar^{j\bdot}.
\ee
On the other hand, since the spinors of different chirality are independent from each other and since  supersymmetry is
manifest, it is clear that one should also consider the complex scalar field $\f$ and its hermitean conjugate $\ft$
defined in Minkowski space as independent fields from each other in Euclidean space. Indeed, this somehow unusual
treatmet appears naturally if one defines a continuous Wick rotations to Euclidean space \cite{vnw}:  pseudoscalar field
$B$ where $\f =A+iB$, goes over into $iB_E$ in Euclidean space i.e. Euclidean scalar fields become  $\f_E=A_E + B_E$ and
$\ft_E=A_E - B_E$. By applying above mentioned continuous Wick rotation  to N=2 SYM
theory in Minkowski space one gets \cite{vnw} the N=2 supersymmetric Euclidean theory that is constructed by Zumino
\cite{zumino}. Note that corresponding action in Euclidean space is hermitean.

Following above remarks, we perform a (continuous) Wick rotation to formulate the results of previous section in
Euclidean space i.e. Minkowskian vector quantities $v^\mu=(v^0,\vec{v})$, $\mu=0,1,2,3$ become Euclidean ones $v_\mu
=(\vec{v},iv^0)$, $\mu=1,2,3,4$ and Euclidean sigma matrices are taken as $e_{\mu\a\adot}=(i\vt,1)$ and
${\bar{e}_{\mu}^{\adot,\a}=(-i\vt,1)}$. We will also take the gauge field anti-hermitean rather then hermitean in order
to follow the instanton literature\footnote{We use the  Euclidean conventions of Ref.\cite{dkhm}.} and to avoid
confusion we will denote Euclidean scalar fields as  $M:=\f_E$,  $N:=\ft_E$.

The Euclidean  $\bs$-transformations of the fields now read
\bea
\bs A_{\mu}&=&  D_{\mu}c - \x_i e_{\mu}\labar^i \\
\bs \la_i &=& - \{c,\la_i \} - e_{\mu\nu}\x_i F_{\mu\nu} +\x_i [M ,N ] + \vt_i ^j \x_j.\vD \\
\bs \labar^i &=& - \{ c,\labar^i \}  +i \kok \bar{e}_{\mu}\x^i D_{\mu} N \\
\bs M &=& - [c, M ] + i\kok \x^i\la_i \\
\bs N &=& - [c, N ]\\
\bs \vD &=& - [c, \vD] + \vt_i ^j(\x_j e_\mu D_\mu \labar^i  + i \kok\x^i [\la_j ,N]) \\
\bs c &=& - \frac{1}{2}\{ c,c\} + i\kok\x_i\x^i N
\eea
and  consequently the gauge fermion (28) in Euclidean space is given as
\be
\Psi_E = \frac{1}{\x _k \x^k} Tr \int d^4 x  (\yarim \x ^i  \la_i  [M ,N ] -\yarim \x^i \vt _i ^j \la_j .
\vD - \frac{1}{2} \x^i e_{\mu\nu}\la_i  F_{\mu\nu} - \frac{i\kok}{2} M \x^i e_\mu D_\mu \labar_i ) .
\ee

The N=2 supersymmetric Euclidean action, that is constructed by Zumino \cite{zumino},  can now be written as the
$\bs$-variation of $\Psi_E$,
\bea
I_E&=&\bs\Psi_E\nonumber\\
&=&Tr \int d^4 x (\frac{1}{4} F_{\mu\nu} F_{\mu\nu}
+\frac{1}{8}\e_{\mu\nu\la\rho}F_{\mu\nu}F_{\la\rho} - \la^i D \!\!\!\! / \labar_i  + M D_{\mu} D_{\mu}
N\nonumber\\
&&\qquad\qquad\qquad
-\frac{i\kok}{2}(\la_i [\la^i , N ] +\labar^i [\labar_i , M ]) -\yarim [M,N ]^2 + \yarim\vD . \vD)
\eea
up to the topological term $\e_{\mu\nu\la\rho}F_{\mu\nu}F_{\la\rho}$ and the auxiliary term $\yarim\vD . \vD$, since in
our approach the inclusion of the auxiliary fields i.e. off-shell formulation is mandatory.

As noted before, the only way to write the action as a $\bs -$exact term is to allow the gauge fermion to depend on the
negative powers of SUSY ghost $\x_i$. In other words $\bs$ has to be defined on the space of field polynomials that are
not necessarily analytic in these constant ghosts.

On the other hand, the above mentioned non-analyticity argument can be used to derive a relation between the above
expressions and topological theory. Note that after twisting physical nature of some fields are interprated differently,
i.e. some fields become ghosts while some others become anti-ghosts \cite{witten,top}.  We summarize the dimensions
and ghost numbers of the topological fields $(A_{\mu},\, \Phi, \, \bar{\Phi},  \,
\psi_{\mu}, \, \eta, \,  \X_{\mu\nu}, B_{\mu\nu})$ in Table2.
%
%
\begin{table}[hbt]
\centering
\begin{tabular}{|c||c|c|c|c|c|c|c|c|}
\hline
&$A_\mu$&$\Phi$&$\bar{\Phi}$&$\psi_{\mu}$&$\eta$&$\X_{\mu\nu}$&$B_{\mu\nu}$&$\,\,c\,\,$\\
\hline
$Dim$&1&0&2&1&2&2&2&0 \\
\hline
$Gh$&0&2&-2&1&-1&-1&0&1 \\
\hline
$GP$&0&0&0&1&1&1&0&1 \\
\hline
\end{tabular}
\caption[t1]{Dimensions   $d$, Grassmann parity $GP$, ghost number $Gh$}
\end{table}

In order to get the topological fields that are given in Table2 that have the correct dimensions and ghost numbers we
note that  the SUSY ghosts $\x_i$ have ghost number one and dimension 1/2. Therefore it is natural to think that the
topological fields can be written as certain functions of ordinary N=2 fields and SUSY ghosts which are also partly
non-analytic in these constant ghosts. These relations can be found by using the non-analytic structure of the gauge
fermion $\Psi_E$ given in (40).  The only consistent field redefinitions that assign the correct dimensionality and
ghost number to the topological fields are found to be\footnote{We have chosen the coefficients in the
definitions of the topological fields in order to get the conventions of \cite{top}.},
\be
A_\mu = A_\mu
\ee
\be
\psi_\mu = - \x_i e_\mu \labar^i
\ee
\be
\Phi=i\kok \x_i\x^i N \quad ,\quad \bar{\Phi}= \frac{i}{\kok\x_i \x^i} M
\ee
\be
\eta = \frac{1}{\x _k \x^k} \x_i\la^i \quad , \quad \X_{\mu\nu}= \frac{-2}{\x _k \x^k} \x^i e_{\mu\nu} \la_i
\ee
\be
B_{\mu\nu}= \frac{-2}{\x _k \x^k} \x^i e_{\mu\nu}\vt _i ^j \x_j . \vD
\ee

It is straightforward to show that when the above variable redefinitions are inserted in the ordinary Euclidean
action (41) and in the transformations (33-39), the action and corresponding BRST transformations that are found, are
exactly the TYM action of Witten \cite{witten} with an auxiliary term and the (extended) BRST transformations
defined in TYM which contain the Witten's scalar SUSY . In other words, as mentioned by several authors (but not shown
explicitly to our best knowledge), TYM theory in flat Euclidean space can be obtained directly as variable redefinitions
from the ordinary N=2 SYM theory. As it is obvious from the above definitions of the topological fields, the ghost
numbers and the dimensions that are assigned to the fields in the twisting procedure  by hand, appears here naturally
due to the composite structure of the topological fields in terms of global ghosts $\x_i$ and the original fields i.e.
with respect to the power of $\x_i$'s in the definitions.

To be clear, (and hoping not to be too tedious) we demonstrate the above points explicitly. By using the field
redefinitions (33-39), $\bs$-transformations can be rewritten as\footnote{Here we note that the derivation of the above
results depends crucially on the commuting nature of the global ghost
$\x_i$. For example it is easy to verify that $\bs$-transformation of $\la_i$ (34) decomposes into (52,53) by using
$\x^i e_{ \mu\nu}\x_i=0$ and $\x^i\vt_i ^j\x_j =0$.}
\bea
\bs A_{\mu}&=& D_{\mu}c + \Psi_{\mu} \\
\bs \psi_\mu  &=& - \{ c, \Psi_\mu \}  -  D_{\mu} \Phi \\
\bs \Phi &=& - [c,\Phi ]\\
\bs c &=& - \frac{1}{2}\{ c,c\} +\Phi
\eea
and
\bea
\bs \bar{\Phi} &=& - [c,\bar{\Phi} ] + \eta \\
\bs \eta &=& - \{c, \eta \} + [\Phi ,\bar{\Phi} ]\\
\bs \X_{\mu\nu} &=& -[c,\X_{\mu\nu} ] + F_{\mu\nu}^+ + \B_{\mu\nu}\\
\bs B_{\mu\nu} &=& - [c, B_{\mu\nu} ] + [\Phi , \X_{\mu\nu} ] - (D_\mu \psi_\nu -D_\nu \psi_\mu)^+
\eea
where $F_{\mu\nu}^+ = F_{\mu\nu}+\frac{1}{2}\e_{\mu\nu\la\rho}F_{\mu\nu}$ is the self-dual part of the field strength
$F_{\mu\nu}$. If one decomposes $\bs$ on the fields $(A_{\mu},\, \Phi, \,\bar{\Phi},  \,\psi_{\mu}, \,\eta, \,
\X_{\mu\nu})$ as
\be
\bs=s_o + \d
\ee
where $s_0$ is the ordinary BRST transformation, one can see that  $\d$-transformations are exactly  the same with the
scalar supersymmetry transformations introduced by Witten \cite{witten}. Note  that this scalar SUSY generator can
also be written as a composite generator,
$$\d=-i\x^i Q_ i $$ where $Q_i$ are the chiral SUSY generators.

In terms of topological fields given in (42-46), the gauge fermion (40) now reads as
\be
\Psi_{top} = Tr \int d^4 x  (-\yarim \eta  [\Phi ,\bar{\Phi} ] + \frac{1}{8}\X_{\mu\nu}  F_{\mu\nu}^+
-\frac{1}{8}\X_{\mu\nu}B_{\mu\nu} + \bar{\Phi} D_\mu \psi_\mu ) .
\ee
and the corresponding action is found as
\bea
I_{top} &=& \bs \Psi_{top}\\
&=&Tr\int d^4 x (\frac{1}{8} F_{\mu\nu}^+ F_{\mu\nu}^+ + \eta D_\mu \psi_\mu -\frac{1}{4} \X_{\mu\nu}(D_\mu \psi_\nu -
D_\nu \psi_\mu)^+ -\bar{\Phi}D^2 \Phi\nonumber\\
 &&\qquad\qquad - \,\,
 \! \yarim \Phi \{\eta ,\eta \} -\frac{1}{8} \Phi \{ \X_{\mu\nu}, \X_{\mu\nu}\} + \bar{\Phi} \{\psi_\mu ,
\psi_\mu  \} -\yarim [\Phi ,\bar{\Phi}]^2 -\frac{1}{8} B_{\mu\nu}B_{\mu\nu} ) .
\eea

The above given action $I_{top}$,  is exactly the Topological Yang Mills action \cite{witten,top} with an auxiliary
field term. We remark once more that the inclusion of the auxiliary field is crucial in order to write the action as an
exact term\footnote{The reason why the action could not be written as an exact term in the original  paper \cite{witten}
is that the twisted theory was obtained from  the on-shell SYM. Note that, since $\Psi_{top}$ given in (56) is gauge
invariant we have $I_{top}=\bs\Psi_{top}=\d\Psi _{top}$.}.

We should stress here that above results do not mean that N=2 SYM is just a topological theory. It  has its own
physical degrees of freedom. However, it becomes a topological theory (of cohomological type) if the analyticity
requirement of the SUSY ghosts is relaxed. This fact has also been pointed out in Ref.s\cite{sorlec,sor1,sor2},
for twisted N=2 SYM that the twist of the N=2 theory can be interpreted as a topological theory only if the analyticity
is lost in (scalar) SUSY ghosts. Note that, in many cases the cohomology of $s$ can be understood by studying a simpler
operator  that is found by using a suitable filtration of $s$ \cite{psbook}. In our case we take $\bs$ as the filtered
operator. The cohomology of $\bs$ is empty only if when $\bs$ is allowed to act on the field polynomials that are not
necessarily analytic in the parameters $\x_i$. Since the cohomology of complete operator $s$ is isomorphic to a subspace
of the filtrated operator $\bs$ \cite{psbook}, the cohomology of $s$ is also empty when the analyticity requirement is
relaxed and the theory can be interprated as a topological theory.

\section{Instantons and Baulieu-Singer approach to TYM}

As it is well known, instantons are finite action solutions of the Euclidean field theories. Aiming to incorporate the
instantons into supersymmetric theories Zumino have constructed a supersymmetric field theory directly in the Euclidean
space \cite{zumino}. This theory has N=2 supersymmetry with a hermitean action. (Recently, this theory has  been
derived by defining a continuous Wick rotation \cite{vnw} and also by using dimensional reduction via time direction
from six dimensional N=1SYM\cite{bvn}.)   It is  observed by Zumino that when one imposes for instance an anti self-dual
field strength,
\be
F_{\mu\nu}^+ = F_{\mu\nu}+\frac{1}{2}\e_{\mu\nu\la\rho}F_{\mu\nu} = 0
\ee
with the following restrictions
\be
 M=\la_i=0
\ee
the equations of motion from (41) reduce to a simple form \cite{zumino},
\bea
F_{\mu\nu}^+ &=& F_{\mu\nu}+\frac{1}{2}\e_{\mu\nu\la\rho}F_{\mu\nu} = 0\\
D^2 N &=& \frac{i\kok}{2}\{ \labar^i , \labar_i \} \\
e_\mu D_\mu \labar^i &=& 0 .
\eea
These restrictions (59,60) and the equations (61-63) are covariant under the supersymmetry transformations found by
applying the above restrictions \cite{zumino}.

The equations (61-63) are also the ones whose solutions are used as approximate solutions of the saddle point
equations in  the context of constraint instanton method \cite{dkm,dkhm}. On the other hand, similar equations are
obtained in TYM without any approximation \cite{bftt2} from an action functional that can be written as a BRST
transformation of a  gauge fermion given by Baulieu-Singer \cite{bs}. Both of the approaches to the instanton
calculations give the same result \cite{bftt2}.

Therefore, since Euclidean N=2 SYM action can be written as a BRST exact term (41) and  Wittens TYM \cite{witten} can be
obtained by using simple field redefinitions (42-46), we look for another analogy between the above instanton
inspired truncation of Euclidean N=2 SYM theory and the Baulieu-Singer approach to TYM.

The first step towards for this purpose is to define a truncated $\ts$ ,
\be
\ts = \bs|_{F_{\mu\nu}^+ = \Ds \labar^i =M = \la_i = 0}
\ee
such that,
\bea
\ts A_{\mu}&=&  D_{\mu}c - \x_i e_{\mu}\labar^i \\
\ts \labar^i &=& - \{ c,\labar^i \}  +i \kok \bar{e}_{\mu}\x^i D_{\mu} N \\
\ts N &=& - [c, N ]\\
\ts c &=& - \frac{1}{2}\{ c,c\} + i\kok\x_i\x^i N
\eea
and
\bea
\ts M &=&  i\kok \x^i\la_i \\
\ts \la_i &=&  \vt_i ^j \x_j.\vD \\
\ts \vD &=& 0
\eea
The reason why we do not set $\la_i = \vD =0$ in equations (69,70) is that the pairs $(M,\x^i\la_i)$ and $(\x^i
\vt_i^j\la_j,\vD)$ behaves like the trivial pairs (or sometimes called BRST doublets) i.e. like $(\cbar,b)$ such that
$s \cbar = b \, ,\, s b=0$ where $s$ is a nilpotent operator. It is known that the cohomology of an operator does not
depend on inclusion of such trivial pairs (see for instance \cite{gps,psbook}).

It is straightforward to derive that $\ts$ is also nilpotent $$\ts^2 = 0$$ and after performing the field redefinition
given in (42-46), $\ts$-transformations are found to be exactly that of Baulieu-Singer \cite{bs, top}.

On the other hand the gauge fermion that is compatible with the restrictions of Zumino \cite{zumino} has to be chosen
slightly different then the one given for Euclidean case (40),
\be
\Psi_{inst.} = \frac{1}{\x _k \x^k} Tr \int d^4 x (  - \frac{\a}{2} \x^i \vt _i ^j \la_j . \vD  - \frac{1}{2} \x^i
e_{\mu\nu}\la_i F_{\mu\nu}^+  + \frac{i\kok}{2}  \x^i e_\mu  \labar_i D_\mu M ) .
\ee

The reason of this modification becomes apparent when the corresponding action is driven,
\bea
I_{inst.}^{(\a)}&=&\ts\Psi_{inst.}\nonumber\\
&=&Tr \int d^4 x (-\frac{\a}{8}B_{\mu\nu}B_{\mu\nu} +\frac{1}{4}B_{\mu\nu}  F_{\mu\nu}^+  - \la^i e_\mu D_\mu \labar_i
 + M (D_{\mu} D_{\mu}N -\frac{i\kok}{2}\{\labar^i ,\labar_i \}) \nonumber\\
&&\qquad\qquad\qquad
 +  \frac{1}{\x _k \x^k} (- \frac{1}{2}\x^i e_{\mu\nu}\la_i [c,F_{\mu\nu}^+] + \frac{i\kok}{2} M \{ c ,\x^i e_\mu D_\mu
\labar_i \}) \,\,) \nonumber\\
&& \qquad\qquad\qquad\qquad\qquad
+  \frac{1}{\x _k \x^k} Tr\int d^4 x \del_\mu( \ts \frac{i\kok}{2}M \x^i e_\mu  \labar_i )
\eea
where we have used the definition of $B_{\mu\nu}$ in order to have notational simplification.

First of all, the gauge fermion $\Psi_{inst}$ (72) and the above action $I_{inst}$ are exactly the ones given in
Baulieu-Singer approach \cite{bs} up to ordinary gauge fixing\footnote{For instance adding $\ts \int d^4 x \cbar
\del^\mu A_\mu)$ to $\Psi_{inst}$ such that $\ts \cbar =b \, ,\, \ts b=0$.}. It is straightforward to rewrite the above
given gauge fermion $\Psi_{inst.}$ and the action $I_{inst.}$ by using the field redefinitions given in (42-46) in order
to get the results of \cite{bs}. However, if the above relations are considered on their own, to be able to derive the
instanton equations (61-63) from the action functional without having any dependence on the constant ghosts, the
Euclidean $\Psi_E$ has to be modified. For instance when  the restrictions of \cite{zumino} are used the $\x_i $
dependence that comes from the $\ts$ variation of $Tr \x^i\la_i [M,N]$ cannot be eliminated. Therefore,  the coefficient
of this term has to be chosen to vanish. The coefficient of $Tr \ts \x^i \vt _i ^j \la_j . \vD$ can be left arbitrary
since after performing the Gaussian integration over the auxiliary field $B_{\mu\nu}$ the action is
\bea
I_{inst.}^{(\a)}&=&\ts\Psi_{inst.}\nonumber\\
&=&Tr \int d^4 x (\frac{1}{8\a}F_{\mu\nu}^+  F_{\mu\nu}^+  - \la^i e_\mu D_\mu \labar_i
+ M (D_{\mu} D_{\mu}N -\frac{i\kok}{2}\{\labar^i ,\labar_i \}) \nonumber\\
&& \qquad\qquad
+  \frac{1}{\x _k \x^k} (- \frac{1}{2}\x^i e_{\mu\nu}\la_i [c,F_{\mu\nu}^+ ] + \frac{i\kok}{2} M \{ c ,\x^i e_\mu
D_\mu \labar_i \}) \,\,) \nonumber\\
&& \qquad\qquad\qquad\qquad
+  \frac{1}{\x _k \x^k} Tr\int d^4 x \del_\mu( \ts \frac{i\kok}{2}M \x^i e_\mu  \labar_i )
\eea
and Fadeev-Popov ghost field, $c$, independent part of the action is also SUSY ghost $\x_i$ free. The
form of the last term in $\Psi_{inst}$ is inspired from Ref.\cite{bftt2} in order to get a surface contribution, since
 $$ Tr\ts \x^i e_\mu  \labar_i D_\mu M  = Tr \del^\mu \ts \x^i e_\mu  \labar_i  M - Tr  \ts M \x^i e_\mu D_\mu  \labar_i
$$ if the scalar field has non-trivial boundary conditions. Therefore, the gauge fermion $\Psi_{inst}$ is the only
consistent choice up to total derivatives that gives the right action to derive the exact instanton equations, when the
truncated transformations (65-71) are used.

On the other hand, the free parameter $\a$ can be thought as a gauge parameter, since
it is so in Baulieu-Singer approach \cite{bs}. By choosing directly $\a=0$, the action (71) takes the form,
\bea
I_{inst.}&=&\ts\Psi_{inst.}\nonumber\\
&=&Tr \int d^4 x (\frac{1}{4}B_{\mu\nu}  F_{\mu\nu}^+  - \la^i e_\mu D_\mu \labar_i
+ M (D_{\mu} D_{\mu}N -\frac{i\kok}{2}\{\labar^i ,\labar_i \}) \nonumber\\
&& \qquad\qquad
+  \frac{1}{\x _k \x^k} (- \frac{1}{2}\x^i e_{\mu\nu}\la_i [c,F_{\mu\nu}^+ ]+ \frac{i\kok}{2} M \{ c ,\x^i e_\mu
D_\mu
\labar_i \}) \,\,) \nonumber\\
&& \qquad\qquad\qquad\qquad
 +  \frac{1}{\x _k \x^k} Tr\int d^4 x \del_\mu( \ts \frac{i\kok}{2}M \x^i e_\mu  \labar_i ).
\eea
and by performing a functional integration over the fields $\la_i \, ,M$ and $\B_{\mu\nu}$, the configurations of the
constraint instanton method (61-63) are obtained without using any approximation procedure. In other words, as it is
demonstrated above, Baulieu-Singer approach can also be obtained by using field redefinitions given in (42-46) when an
instanton inspired truncation (59,60) of the Euclidean model is used.

\section{Conclusion and discussion}

In this note, we have shown how to write the off-shell N=2 SYM action as an exact term by using a nilpotent
extended BRST operator $\bs$ that includes supersymmetry transformations of one chirality. The corresponding gauge
fixing fermion is found to be nonanalytic in global SUSY ghosts. In other words it is shown that the action
belongs to trivial cohomology of the extended BRST operator $\bs$, if this operator is allowed to act on the field
polynomials that are not necessarily analytic in these global ghosts.

Due to this nonanalytical structure, we have found field redefinitions such that Witten's TYM theory \cite{witten}
can be obtained from the Euclidean N=2 SYM by identifying the fields of TYM with composite fields of N=2
vector multiplet and the chiral SUSY ghosts $\x_i$ in Euclidean space. These field redefinitions are also partly
non-analytic in global SUSY ghosts. The ghost numbers and the dimensions of the topological fields, that are assigned by
hand when it is formulated by twisting, appear in our approach naturally according to this composite structure. In other
words, we have shown explicitly that TYM theory can be found from N=2 SYM exactly as a change of variables (i.e without
twisting).

The above mentioned analyticity requirement also plays a decisive role to understand when N=2 SYM can be
interprated  as a topological theory. Note that, the topological theory is obtained via change of variables only when
the requirement of analyticity of the constant ghosts $\x_i$  is relaxed. Therefore  physical and topological
interpretations of N=2 SYM are intertwined together. However, in order to have a better understanding of implications of
the results presented above, it would be intresting to investigate the perturbative regime of the theory  by using
the standard techniques of algebraic renormalization framework \cite{psbook}.

On the other hand, when the restrictions on the fields in order to get supersymmetric instanton
configurations  \cite{zumino} are used, we show that with the help of a  truncated BRST operator $\ts$, an
action can be written as an $\ts$-exact term. After using the given variable redefinitions, it is seen that this
formulation is  exactly  TYM theory in the approach of Baulieu-Singer \cite{bs}. The instanton equations, that
are used in the {\it traditional} instanton computations \cite{dkm,dkhm} can be derived from this action (73) without
using any approximation. Moreover, it is known that the Witten's action \cite{witten}  can be obtained from the one
given in Baulieu-Singer approach by a continuous deformation of the gauge fixing \cite{bs}. As a consequence a similar
relation also occurs between Euclidean N=2 SYM action (41) and the truncated (instanton) action (73) since both of
the formulations of TYM can be obtained by using the variable redefinitions.

Finally, it is worthwhile to mention that the instanton calculations performed in Baulieu-Singer approach
of TYM \cite{bftt2}  gives exactly the same result with the one performed in N=2 SYM  \cite{dkm, dkhm}.   Since both N=2
SYM and TYM theories are shown to be equivalent by simple variable redefinitions, it would also be intresting to
reinvestigate the equivalence of the instanton calculations of N=2 SYM and TYM and to find out whether the instantons
localize in the topological sector of the theory where the functional space of field polynomials is not necessarily
analytic in global SUSY ghosts or not.

\vskip 2cm
{\it Acknowledgments

\noindent I am indebted to R. Flume for an initiation to the subject. \\
I also gratefully acknowledge the  numerous enlightening discussions with \"{O}. F. Day\i{} and M.
Horta\c{c}su.}

%
%
%
%

\end{document}